\documentclass[12pt,preprint]{aastex}
\usepackage{natbib}
\newcommand{\lya}{Lyman-$\alpha$}

\begin{document}

\title{Formation of Metal-Poor Globular Clusters in Lyman $\alpha$ Emitting Galaxies
in the Early Universe}

\author{Bruce G. Elmegreen}
\affil{IBM Research Division, T.J. Watson Research Center, 1101 Kitchawan
Road, Yorktown Heights, NY 10598, USA, bge@watson.ibm.com}

\author{Sangeeta Malhotra and James Rhoads}
\affil{School of Earth and Space Exploration, Arizona State University,
Tempe, AZ 85287, USA}

\begin{abstract}
The size, mass, luminosity, and space density of Lyman-$\alpha$ emitting
(LAE) galaxies observed at intermediate to high redshift agree with
expectations for the properties of galaxies that formed metal-poor halo
globular clusters (GCs).  The low metallicity of these clusters is the
result of their formation in low-mass galaxies. Metal-poor GCs could enter
spiral galaxies along with their dwarf galaxy hosts, unlike metal-rich GCs
which form in the spirals themselves. Considering an initial GC mass larger
than the current mass to account for multiple stellar populations, and
considering the additional clusters that are likely to form with massive
clusters, we estimate that each GC with a mass today greater than
$2\times10^5\;M_\odot$ was likely to have formed among a total stellar mass
$\gtrsim3\times10^7\;M_\odot$, a molecular mass $\gtrsim10^9\;M_\odot$, and
$10^7$ to $10^9\;M_\odot$ of older stars, depending on the relative gas
fraction. The star formation rate would have been several $M_\odot$
yr$^{-1}$ lasting for $\sim10^7$ yrs, and the Lyman-$\alpha$ luminosity
would have been $\gtrsim10^{42}$ erg s$^{-1}$. Integrating the LAE galaxy
luminosity function above this minimum, considering the average escape
probability for Ly$\alpha$ photons (25\%), and then dividing by the
probability that a dwarf galaxy is observed in the LAE phase (0.4\%), we
find agreement between the co-moving space density of LAEs and the average
space density of metal-poor globular clusters today. The local galaxy WLM,
with its early starburst and old GC, could be an LAE remnant that did not
get into a galaxy halo because of its remote location.
\end{abstract}


\section{Introduction}
Metal-poor globular clusters (GCs) that inhabit the halos of spiral galaxies
\citep{brodie06} could have arrived there in dwarf galaxies that got
captured by dynamical friction and dispersed by tidal forces
\citep{searle78, zinnecker88,freeman93}. Many GCs are still in debris
streams \citep{dacosta95, palma02, mackey04, carraro07, casetti09,
newberg09, mackey10} or share orbits with other GCs that presumably came
from the same dwarfs \citep{gao07,smith09}. The low metal abundance of halo
GCs compared with disk and bulge GCs \citep{strader07, alves11} could then
be the result of the mass-metallicity relation for galaxies
\citep{chies-santos11}, which is established early on
\citep{erb06,maiolino08, mannucci09}.  A mass-metallicity relation is also
present for GCs themselves \citep{wehner08,forbes10, mieske10}, although
self-enrichment may be the reason \citep{bailin09}. The most heterogeneous
GCs could have been dwarf galaxy nuclei \citep[e.g.,][]{bekki06,carretta10}.

Metal-poor GCs in elliptical galaxies would have formed in the same types of
dwarfs as metal-poor GCs in spirals. Some of these GCs got into the
ellipticals after two or more spirals merged \citep{ashman92}, while others
got in with captured dwarfs after the ellipticals formed \citep{cote98}.
Debris streams containing GCs are observed in ellipticals too
\citep{romanowsky12}. The larger radial distribution of metal-poor GCs
compared with metal-rich GCs in ellipticals
\citep{harris09,forbes11,liu11,faifer11} presumably results from this
difference in their formation sites.

Metal-rich GCs have a different distribution than metal-poor GCs in spiral
galaxies, suggesting that they formed in the disks and bulges of these
galaxies along with the other disk stars \citep{larson88}, perhaps during a
clumpy phase \citep{shap10} when the thick disk and bulge formed
\citep{beasley02,griffen10,bournaud09}. Their higher metallicities result
from the generally higher metallicities of their more massive hosts.
Metal-rich GCs also get into ellipticals during the mergers of spirals, and
other metal-rich GCs form in ellipticals during the merger itself
\citep{ashman92}. Ellipticals then have a higher specific frequency of
metal-rich GCs than spirals \citep{harris91} because of the additional GCs
that formed during the merger. Ellipticals should also have a higher
specific frequency of metal-poor GCs if their more central locations in
dense galaxy clusters gave them better access to a rich population of
remnant GCs in cluster dwarfs.

The mass distribution functions of metal-poor and metal-rich GCs are about
the same \citep{wehner08}. Both have a peaked shape with the same
characteristic mass ($\sim2\times10^5\;M_\odot$). In the dwarf model for
metal-poor GCs, this similarity requires that the combined mass distribution
of small numbers of clusters from many dwarf galaxies is the same as the
single mass distribution of many clusters in a large galaxy. Observations
confirm this similarity for old massive clusters in local dwarf and spiral
galaxies \citep{kruijssen12}. The mass function of clusters is apparently a
local property of star formation independent of the number of clusters
formed \citep{elm97}. The change from an initial power law or Schechter mass
function to a peaked function is presumably the result of selective
disruption of low mass clusters shortly after birth
\citep[e.g.,][]{parm05,elmegreen10} and a stellar evaporation rate that is a
weak function of cluster mass \citep{mf08}.

Here we consider the formation conditions for GCs in modern environments
(Sect. 2) and assess whether metal-poor GCs are likely to have formed in
galaxies that have already been observed at high redshift (Sect. 3). The
most likely candidates are small active galaxies like Lyman $\alpha$
emitters (LAE). Earlier attempts to link GC formation with specific galaxies
or types of galaxies \citep[e.g.,][]{larson88} lacked the extensive data on
high-redshift star formation that are now available in large surveys. A
comparison with other models is in Section 4.

\section{Conditions for GC Formation in the Local Universe}

In the solar neighborhood of the Milky Way, star clusters form in the midst
of other clusters and unbound stars, with a mass that is at most a few
percent of the associated molecular cloud mass. One of the nearest regions
where star formation has produced a total stellar mass comparable to that in
a globular cluster is Eta Carina, where instead of a single cluster there are
hundreds of clusters up to $\sim10^3\;M_\odot$ in mass and about three times
as much mass in free stars \citep{feigelson11}. A bigger star-forming region
is W31 at a distance of 6 kpc, which emits $6\times10^6\;L_\odot$ of young
star radiation, contains several ultra-compact HII regions and subclumps of
$\sim10^{5.8}\;L_\odot$ each, masers and infrared prestellar clumps. The
dense molecular clumps, however, extend up in mass to only
$\sim8000\;M_\odot$ because they are part of an $dn(M)/dM\propto M^{-1.5}$
power law of clump masses that distributes the total molecular mass of
$1.2\times10^5\;M_\odot$ among smaller pieces (Beuther et al. 2011). The
largest cluster likely to come from such clumps is only several thousand
$M_\odot$.

Looking for even larger regions in the Milky Way, one of the most extreme is
W43, located at the near tip of the main stellar bar. Such bar-end regions
typically house prodigious star formation, presumably because of the massive
collection of gas in orbit around the bar \citep{kenney91}. W43 extends for
310 pc and contains a molecular mass of $7\times10^6\;M_\odot$ in addition
to an atomic mass of $3-6\times10^6\;M_\odot$ (Luong et al. 2011). The
fraction of molecules in dense $^{13}$CO substructures is high, $\sim10$\%,
and the CO line FWHM is large, 22 km s$^{-1}$. This is the type of region
that might produce a $10^5\;M_\odot$ cluster along with many smaller
clusters (Luong et al. 2011). It took the Milky Way bar to make these
conditions, which means a total distance for gas collection and compression
in the bar flow exceeding several kiloparsecs.

For a power law cluster mass function, $dn(M)/dM=n_0M^{-\beta}$, the ratio of
the total cluster mass to the largest cluster mass is
\begin{equation}
{{M_{\rm total}}\over{M_{\rm max}}}=1+{{\int_{M_{\rm min}}^{M_{\rm max}} M^{1-\beta} dM}
\over {M_{\rm max}\int_{M_{\rm max}}^{\infty}M^{-\beta}dM}}
\end{equation}
for minimum cluster mass $M_{\rm min}$. This ratio implies that the minimum
likely young stellar mass to form a single largest cluster of mass $M$ is
\begin{equation}
M_{\rm total}=\eta_{\rm c}^{-1}M+\eta_{\rm c}^{-1}(\beta-1)M^{\beta-1}
\left({{M^{2-\beta}-M_{\rm min}^{2-\beta}} \over {2-\beta}},
\ln\left[{{M}\over{M_{\rm min}}}\right]\right),\label{eq2}
\end{equation}
where $\eta_{\rm c}$ is the fraction of young stellar mass that forms in
bound clusters as opposed to the field, and the second result is for
$\beta=2$. Observations of Eta Carina by \cite{feigelson11} suggest
$\eta\sim 0.25$. For typical $\beta\sim2$ and minimum cluster mass $M_{\rm
min}\sim10\;M_{\odot}$, the minimum total young stellar mass to form a
$10^6\;M_\odot$ cluster is $5\times10^7\;M_\odot$. With an average star
formation efficiency in molecular clouds of $\sim2$\%, the associated
molecular mass has to exceed $2.5\times10^9\;M_\odot$.  We view this as the
necessary molecular mass in a whole galaxy, since by statistical arguments,
subclumps that form individual clusters need not all occur in the same
shielding envelope for $H_2$ formation.  Similar discussions leading to a
large total molecular mass for GC formation were made by \cite{larson88} and
\cite{harris94}.

For typical gas fractions of 50\% or more at high redshift, as determined
from CO observations or star formation rates converted to gas mass using the
\cite{kennicutt98} relation \citep{mannucci09,daddi10,tacconi10}, this
result for gas mass implies that the stellar mass in a galaxy that forms a
single $\sim10^6\;M_\odot$ GC plus all of the other clusters and bare stars
that accompany it in the starburst is typically around $10^9\;M_\odot$. For
90\% gas fractions at the low metallicities of GCs \citep[Fig. 8
in][]{mannucci09}, the underlying stellar mass might be $\sim10^8\;M_\odot$.
Lower mass clusters would generally be associated with lower masses of
peripheral stars, and the peripheral mass would be lower still if the GC
forms in the first galaxy starburst (100\% gas fraction).

An upper mass cutoff for clusters giving a Schechter distribution function
for mass \citep[i.e., a power law with an exponential tail --][]
{schechter76} instead of a power law has been suggested for spiral galaxies
by \cite{gieles06a,gieles06b} and \cite{larsen09} and for elliptical
galaxies by \cite{waters06} and \cite{jordan07}. \cite{larsen09} measured an
average cutoff (where the exponential tail begins) for several spirals at
$\sim2\times10^5\;M_{\odot}$. Such a cutoff can increase the required cloud
mass if the maximum cluster mass in a distribution is larger than the cutoff
mass. Cutoffs have not been explained theoretically except for a possible
connection with interstellar pressure \citep{elmegreen01}. The application
of such a cutoff to the present problem is uncertain because massive young
clusters with $M\sim10^6\;M_{\odot}$ are observed in dwarf galaxies like NGC
1569 \citep{ho96} and NGC 1705 \citep{meurer95}; there is no observation of
a statistically significant cutoff in dwarfs.

The star formation rate for a molecular mass of $2\times10^9\;M_\odot$ is
$\sim1\;M_\odot$ yr$^{-1}$ for a conventional molecular consumption time of
2 Gyrs \citep{bigiel08,leroy08}. This is the characteristic rate for a
spiral galaxy today, and large for a dwarf galaxy. If the molecular
consumption time was shorter by a factor of $\sim4$ in the early universe
\citep{genzel10}, then a dwarf at that time would have had a very intense
starburst, possibly enough to disturb the ISM to a significant degree
\citep[e.g.][]{dekel86,Finkelstein11a,mclinden11}.

We are interested here in whether such a star-forming dwarf could be
observed at about the time when it forms a $10^6\;M_\odot$ cluster. For a
star formation rate of $4\;M_\odot$ yr$^{-1}$ lasting $\sim100$ Myr, which
is the orbit time of a 0.5 kpc radius disk with a typical dwarf galaxy
rotation speed of 30 km s$^{-1}$, the SDSS g-band absolute AB magnitude
would be $-20.0$ and the young-star luminosity $1\times10^{10}\;L_\odot$
(according to models in Bruzual \& Charlot 2003 for a metallicity of 0.2
solar; emission lines would make the galaxies brighter). For redshifts $z=2$
and $z=6$, where the distance moduli in the observer frame are 46.0 and 48.9
and the restframe g-band shifts to 1.4$\mu$m and 3.3$\mu$m, the apparent AB
magnitudes would be 26.0 and 28.9, respectively. These magnitudes are bright
enough to be detected in deep surveys. Similarly, the absolute g-band AB
magnitude of a model cluster 100 Myr old with $10^6\;M_\odot$ at birth can
be determined from Bruzual \& Charlot (2003) to be $-12.46$. At $z=2$ and 6,
the apparent magnitudes of such a cluster would be 33.5 and 36.4 mag, which
are beyond the current limits.

Also related to observability is the diameter of a typical dwarf galaxy,
$\sim1$ kpc, which corresponds to an angular size of 0.12 arcsec at $z=2$
and 0.17 arcsec at $z=6$. These sizes can be resolved by HST and some
ground-based instruments with adaptive optics, which implies that the
formation environments for metal-poor GCs can be observed directly. The
$10^6\;M_\odot$ clusters themselves would not be resolved, however.

The co-moving space density of dwarf galaxies that formed GCs during an
intense starburst should have been about twice the space density of
metal-poor GCs today, considering one forming GC per dwarf and an
evaporative loss of about a factor of 2 during the intervening time
\citep{ves98}. The space density of all GCs today is $\sim8$ Mpc$^{-3}$
\citep{pm00}, of which about half to two-thirds are metal-poor
\citep{forbes00}, i.e., the blue GCs in the
bimodal color distribution.  This makes the expected dwarf galaxy co-moving
density at the time of metal-poor GC formation equal to about $8$
Mpc$^{-3}$, although only one quarter of these produce GCs more massive than
the peak in today's GC mass function. We discuss these densities in more
detail below.

The metallicities of metal-poor GCs should be about the same as the gas
metallicities of the galaxies they formed in, which is low for low-mass
galaxies at intermediate to high redshift. \cite{zinn85} considered that
metal poor halo GCs have $[\rm {Fe/H}]<-0.8$ and \cite{muratov10} considered
$[\rm {Fe/H}]<-1$. If we consider the solar abundance to be $12+\log(\rm
{O/H})=8.66$ \citep{asplund04}, then metal-poor GCs have $12+\log(\rm
{O/H})<7.7$ or 7.8.  This value is consistent with the observed gas
metallicities of galaxies with stellar masses less than $10^9\;M_\odot$ and
redshifts greater than $z\sim3.5$ \citep{maiolino08,mannucci09}.

The local galaxy WLM, at a distance of $\sim1$ Mpc \citep{mcconnachie05}, is
an example of a dwarf with a metal-poor GC and a very early starburst of
about the same age. According to \cite{dolphin00}, the galaxy has an iron
abundance that rises from [Fe/H]$=-2.18\pm0.28$ to $-1.34\pm0.14$ in a
prolonged starburst that occurred 12 Gyr to 9 Gyr ago, and it has a globular
cluster \citep{ables77} with an abundance of [Fe/H]$=-1.63\pm0.14$
\citep[from][]{hodge99} that also has a very old age, estimated to be
$\sim14$ Gyr by \cite{hodge99}.  Enhanced $\alpha$ elements in the GC
suggest a period of rapid star formation in the gas from which it formed
\citep{colucci11}. Most of the WLM halo stars are also very old
\citep{minniti97}. The stellar mass of WLM today is
$\sim1.6\times10^7\;M_\odot$ \citep{zhang12}; the total dynamical mass
within the half-light radius of 1.6 kpc is $4.3\pm0.3\times10^8\;M_\odot$
\citep{leaman12}.  The V-band absolute magnitude of the GC is $-8.8$ mag
\citep{sandage85,hodge99}, which gives it a luminosity 4.8 times greater
than the luminosity at the peak of the Milky Way globular cluster
distribution function.  Scaling the GC mass accordingly makes it
$10^6\;M_\odot$, or 10\% of the current galaxy stellar mass. \cite{leaman12}
suggest that the WLM stellar mass was only $8\times10^5\;M_\odot$ 11 Gyr
ago, which is comparable to the GC mass at the same time. Evidently, the GC
in WLM was the dominant part of a major starburst shortly after the galaxy
formed. It remains inside WLM and not in the halo of the Milky Way or M31
because of its large distance from both, $\sim1$ Mpc \citep{leaman12}.  We
propose that most of the other metal-poor GCs that formed in young dwarfs in
the local group got captured by the Milky Way, M31, or M33, or possibly also
by the LMC and SMC, before the rest of the dwarf galaxy gas had a chance to
form more stars like WLM did.  Thus the remnant streams may have low stellar
masses like WLM did when it formed its GC; this is much less than the
stellar mass in WLM today.

In summary, a wide variety of evidence points to the formation of metal-poor
globular clusters in low mass galaxies at intermediate to high redshift.
Simple considerations suggest that such galaxies could be observed with
modern instruments during their GC-formation phase, although the young GCs
themselves probably cannot be distinguished. We ask now whether the active
phases of the host galaxies have been observed already.

\section{Lyman $\alpha$ Emitters as Sites for GC Formation}

Small, metal-poor galaxies at high redshift are a natural choice for the
birth place of metal-poor globular-clusters. \lya\ emitting galaxies are
typically dwarf star-forming galaxies at intermediate to high redshift. Most
LAE galaxies have low stellar masses, typically between $10^7$ and $10^8\;
M_\odot$ \citep{Pirzkal07,Finkelstein07}, although a small fraction have
higher mass following a Schechter luminosity function
\citep{Lai07,Finkelstein08,Finkelstein09a}. The Lyman $\alpha$ emitting
regions in LAE galaxies are usually very young, with stellar ages of a few
$10^7$ yrs
\citep{Pirzkal07,Gawiser07,Finkelstein07,Finkelstein08,Finkelstein09a}.
Sometimes there is an older underlying population of stars too
\citep{nilsson11,acquaviva12}. LAEs are usually compact
\citep[$\sim$kpc;][]{Pentericci09,Bond09,Bond10,finkelstein11x}, with a
characteristic radius that does not evolve much with redshift
\citep{Malhotra12}, although the LAEs with underlying populations could be
growing in mass \citep{nilsson11}. This paradigm explains the number
densities of LAE galaxies as well as their clustering statistics
\citep{Kovac07,Gawiser07,Tilvi09}. LAEs are also seen to be metal poor
\citep[$\sim0.1$ solar;][]{Finkelstein11a,Richardson12}.

LAE galaxy properties can be explained by postulating that most of them are
dwarf galaxies emitting \lya\ in the first $\sim 3 \times 10^7$ years after
a major starburst \citep{Malhotra02}. The escape of Lyman $\alpha$ radiation
would favor a porous interstellar medium in a small galaxy, so there may be
a selection effect for dwarfs among LAE samples.  This selection works in
our favor because we are looking for dwarfs that had starbursts at
intermediate to high redshift. We postulate that these dwarfs are the
formation sites of metal-poor GCs, and the precursors of the remaining
stellar streams that accompanied these GCs into galaxy halos. Extreme
emission line objects at intermediate redshift are in the same category, and
with their enormous starbursts, could have made most of today's dwarfs
\citep{wel11}.

Can LAEs produce most of the metal-poor GCs seen at the present time? The
mass function of GCs today is a log-normal distribution centered at $\sim 2
\times 10^5\;M_\odot$ \citep{mc03}. These clusters have lost some of their
original mass by evaporation.  \cite{mf08} fit the current GC mass function
with an initial Schechter function and an evaporated mass per cluster of
$\Delta=1.45\times10^4\;M_\odot(\rho_h/M_\odot\;{\rm pc}^{-3})^{1/2}$ for
cluster density $\rho_h\sim10-1000\;M_\odot$ pc$^{-3}$ inside the half-light
radius. For average $\rho_h=246\;M_\odot$ pc$^{-3}$ in the Milky Way,
$\Delta=2.3\times10^5\;M_\odot$ \citep{mf08}. The initial GC mass would have
been at least this much larger than its current mass.

We estimated above that the number density of original metal-poor clusters
had to be about $\sim 8$ Mpc$^{-3}$. This came from a current density of
$\sim4$ Mpc$^{-3}$, multiplied by 2 to account for evaporation. For a nearly
constant evaporation rate per cluster, this estimate of initial density
should be done again, now in two parts. First, we consider all clusters with
a mass today larger than $2\times10^5\;M_\odot$, which is both the mass at
the peak of the GC mass function and the total evaporated mass per cluster,
on average. An approximation is that all clusters originally larger than
this are still present today because they have not completely evaporated
yet. Their density is about half the current density of metal-poor GCs,
giving $\sim2$ Mpc$^{-3}$, because they represent half of the current GC
mass function by number.  Thus we ask whether galaxies going through the LAE
phase and forming clusters more massive than $2\times10^5\;M_\odot$ today
had a co-moving space density of around 2 Mpc$^{-3}$.

Second, we consider lower mass clusters. These are on the low-mass,
decreasing part of the current GC mass function and should be highly
evaporated from their initial state.  The mass function of clusters at birth
is usually observed to be $dn(M)/dM\propto M^{-2}$
\citep[e.g.,][]{degrijs06,larsen09}. If we consider as representative the
lower part of the GC mass function down to 1/4 the mass of the peak, then
the initial number of clusters below the peak, i.e., between 1/4 and 1 times
the current peak, is 3 times the number of clusters with masses greater than
the current peak, based on this initial cluster mass function. Presently,
there is an equal number of clusters with a mass below and above the peak,
so 1/3 of those below the peak survive. The total survival fraction is still
the factor of 2 estimated in the introduction, but now we have divided it up
between 100\% survival at $M>2\times10^5\;M_\odot$ and 33\% survival at
initial $M<2\times10^5\;M_\odot$.  (That is, for every GC today at
$M>2\times10^5\;M_\odot$ there was one GC originally, and for every GC today
at $M<2\times10^5\;M_\odot$ there were three GCs originally). We first check
whether clusters above the current peak could have formed in LAE galaxies.

To make a cluster of mass $2 \times 10^5\;M_\odot$ today, the initial
stellar mass had to be larger, not only because of evaporation over a Hubble
time, as discussed above, but also to account for the high abundance of
elements from first-generation stellar ejecta that are present in
second-generation stars. Globular clusters have multiple stellar populations
\citep[see reviews in][]{renzini08, carretta10b}.  Models by
\cite{dercole08,dercole12}, \cite{bekki11}, \cite{schaerer11} and others
suggest that the initial cluster mass was $\sim10^6\;M_\odot$ or more, ten
times larger than the present day mass. These models are uncertain, but to
be consistent with them, we assume that the initial characteristic cluster
had $2\times10^6\;M_\odot$. It would most likely have been accompanied by
other clusters, making the total young stellar mass
$\sim3\times10^7\;M_\odot$ if $\eta=1$ in equation (\ref{eq2}). Most of the
star formation that made these clusters, whether in the first or the second
generations, would presumably occur in $\sim10^7$ years, before supernovae
could remove the star-forming gas \citep{dercole12}. However, the total span
of cluster formation could have taken $\sim100$ Myr, considering the need
for enriched ejecta from first-generation stars to re-accumulate along with
fresh material to form the second generation. Non-clustered stars would most
likely have been present with the first generation too ($\eta<1$), but they
would presumably be relatively dim by the time the second generation stars
formed in the final cluster burst. Considering the many uncertainties that
are still present in the detailed theory of GC formation, we use here a
representative luminosity for the starburst phase that is equivalent to
$3\times10^7\;M_\odot$ of stars forming in $\sim10$ Myr. This gives a
star-formation rate during the main starburst equal to $3\;M_\odot$
yr$^{-1}$.

Another consideration is the absorption by dust of Lyman continuum radiation
before it ionizes the nearby gas and produces Lyman $\alpha$ radiation. We
assume a small amount, $A_{\rm LC}\sim1$ magnitude of extinction for Lyman
continuum radiation \citep[e.g.,][]{finkelstein11c,walter12}. That reduces
the effective star formation rate for Lyman $\alpha$ production to
$\sim1\;M_\odot$ yr$^{-1}$. This implies a \lya\ line luminosity of $1
\times 10^{42} {\rm\,erg\,s^{-1}}$ if the escape fraction of the \lya\
photons is 100\%. Recent studies of LAE galaxies show that the average
escape fraction of \lya\ photons is between 17\% and 30\%
\citep{Blanc11,Zheng12,Richardson12}. Assuming an average escape fraction of
25\%, a SFR of 1 $M_\odot$/year translates into \lya\ line luminosity of
$2.5 \times 10^{41} {\rm\,erg\,s^{-1}}$.

Now we integrate the \lya\ luminosity function \citep{Zheng12b} with
$\log(L_*) = 42.86 \pm 0.06$ and $\phi_*=-3.55\pm 0.09$ down to this minimum
luminosity of $2.5 \times 10^{41} {\rm\,erg\,s^{-1}}$. We assume a slope of
$-2$ at low mass. This gives the number density of \lya\ galaxies brighter
than the cutoff for forming a characteristic GC equal to $7 \times 10^{-3}\;
\hbox{Mpc}^{-3}$.

In deriving this density, we have assumed that the \lya\ phase lasts for
$10^7$ years, whereas we observe that the \lya\ luminosity function is
essentially unchanged from $z=2$ to $7$, a time interval of 2.56 Gyrs. This
duration for the LAE morphology corresponds to $\sim 256$ generations of
individual \lya\ events, or a probability of catching a low-mass galaxy in
the LAE phase equal to 0.39\%. Dividing the observed number density of LAE
galaxies that can make today's $M>2\times10^5\;M_\odot$ GCs by the
probability of observation produces a total number density of such galaxies
that go through the LAE phase equal to $\sim 1.9 \;\hbox{Mpc}^{-3}$. If each
LAE phase forms a single GC with a present-day mass of
$M>2\times10^5\;M_\odot$ (along with smaller clusters, as discussed above),
then this would be the comoving number density of metal-poor GCs that end up
more massive than the median mass of today's GCs.  This is in fact the
initial co-moving density of low-metal, high-mass GCs that we expected from
the above discussion of current GC density and evaporation rate.  Thus dwarf
LAE galaxies at redshifts from 2 to 7 could have formed most of the metal
poor GCs observed today in the high-mass half of the GC mass distribution
function.

Formation of the low-mass GCs requires a higher space density of
lower-luminosity LAEs because there were three time as many initially
low-mass GCs as high mass GCs.  This result is easily obtained, however,
because the LAE luminosity function, presumed equal to the LAE mass
function, has the same slope as the initial GC mass function, namely, $-2$
for equal intervals of mass. Thus there were 3 times as many LAEs between
1/4 and 1 times the fiducial luminosity estimated above as there were above
the fiducial luminosity, just as there were 3 times as many initial GCs
below the current peak as above. We conclude that all of the GCs could have
been made in LAE galaxies, based these assumptions.

The above estimates can be summarized by two equations:
\begin{equation}
n(M > M_{*,min}) = {T\over t_{LyA}} \int_{L_{min}}^\infty \Phi(L) dL
=  {T\over t_{LyA}} \Phi^* \Gamma(\alpha+1,L_{min} / L^*)
\end{equation}
and
\begin{equation}
 L_{min} = {M_{*,min} \over  t_{LyA}} f_{esc,LyA}10^{-0.4A_{\rm LC}}  \times {10^{42}
\hbox{erg}\; \hbox{s}^{-1}
\over M_\odot \hbox{yr}^{-1} },
\end{equation}
where $M_{*,min}\approx13M_{\rm cl}$ from equation (\ref{eq2}) is the
minimum stellar mass formed during the starburst lasting for the time
$t_{LyA}$. We assume a luminosity function described by a Schechter
function. Our calculation above used fiducial values of $T=2.56
\;\hbox{Gyrs}$, $t_{LyA} = 10^7\;\hbox{yr}$, $f_{esc,LyA}\sim0.25$, $A_{\rm
LC}\sim1$, $\log(\Phi^* \times \hbox{cMpc}^3) \approx -3.55$, $\log({L^* /
\hbox{erg}\,\hbox{s}^{-1}}) \approx 42.86$, $\alpha = -2$, and $M_{\rm cl}
\approx 2\times 10^6 M_\odot$.

These equations show that the space density of GCs derived from the LAE
luminosity function is insensitive to the lifetime $ t_{LyA} $ of the \lya\
phase. If we assume that the life span of the \lya\ emitting phase goes up
to $t_{LyA}= 3 \times 10^7$ years, then the minimum SFR and the \lya\
luminosity required would be lower by a factor of three. The number of \lya\
emitters brighter than this cutoff would increase by roughly a factor of 3.
However, the total number of GCs produced over the redshift interval $z=2$
to 7 would remain the same, since our 2.56 Gyr interval is now only $\sim
85$ generations of galaxies.

Formal uncertainties in the luminosity function parameters taken from
\cite{Zheng12b} change the final numbers by only about 6\%, since the errors
in $\phi_*$ and $\L_*$ are anticorrelated and roughly cancel out. There is
however a larger spread in the luminosity function parameter found for LAEs
by different authors and surveys. Taking two extremes: (1) \cite{hu10} found
$\phi_*=-3.96$ and $\L_*=43.0$ at z=5.7, which would yield about half of the
GC number density calculated above; (2) \cite{ciardullo12} found
$\phi_*=-2.96$ and $\L_*=42.6$ at z=3.1 yielding 2.3 times the number
density derived above. A change of slope in the faint end of the luminosity
function from $\alpha=-2.0$ to  $\alpha=-1.8$ reduces the number of GCs to
60\% of the above value. The faint end of the luminosity function is quite
steep, however, in both LAEs \citep{rauch08} and Lyman break galaxies
\citep{yan04}; thus, $-2$ may be a more reasonable estimate for the slope.

The Lyman-$\alpha$ escape fraction is also uncertain, perhaps by a factor of
1.6 to 2, as is the extinction for Lyman Continuum radiation, $A_{\rm LC}$,
the initial representative cluster mass, $M_{\rm cl}$, and the fraction of
stars formed in clusters, $\eta$. A lower escape fraction, higher $A_{\rm
LC}$, lower $M_{\rm cl}$, or higher $\eta$ will all lower the observed
Lyman-$\alpha$ luminosity from a GC-forming galaxy, and this increases the
space density of such galaxies according to the luminosity function. The
resulting density of GCs increases in proportion. Lower $M_{\rm cl}$, as
recently proposed by \cite{larsen12}, can be accompanied by a proportionally
lower cluster formation fraction, $\eta$, and not change the GC space
density much.  The formation of more than one GC per LAE galaxy would not
increase the space density of GCs, because we are accounting already for all
of the luminosity associated with GC formation. If 2 GCs are made in each
LAE, then the LAEs would have twice the fiducial luminosity (or each would
form a single GC twice as often), and about half the estimated number of
LAEs would be forming these GCs. The total number of final GCs remains about
the same.

Other small galaxies in the early universe could be formation sites for GCs
too, but if these galaxies are not Lyman-$\alpha$ emitters, then they are
probably not in the starburst phase at the time of observation, and whatever
GCs they formed came about earlier (or may come about later). In that case,
they are already counted in the above discussion when we considered the
$\sim0.39$\% probability for the LAE phase.  In addition, some dwarf
starburst galaxies that form metal-poor GCs may have low Lyman $\alpha$
escape fractions \citep[e.g.][]{heckman11}, placing them below our limit of
integration over the luminosity function.  These galaxies balance those with
high escape fractions, since we use only the average value of 25\% here as
an estimate for this fraction. More massive galaxies would have formed GCs
in starbursts as well, but then the GC metallicities would be larger at the
redshifts when typical galaxies became massive, and the result would be
metal-rich GCs. This is the likely formation mechanism for metal-rich GCs,
which are still correlated with the disk and bulge populations of spiral
galaxies \citep{shap10}.

\section{Comparison with other Models}

Previous models of GC formation in a cosmological context
\citep[e.g.,][]{bromm02,beasley02,kravstov05,moore06,bekki08,griffen10,muratov10}
contain many aspects of the present proposal without identifying likely
hosts that could be visible at intermediate to high redshift.
\cite{beasley02} discussed metal-poor GC formation in gassy disks at early
times and suggested that the GC formation sites might be damped Ly$\alpha$
systems, which are gas clouds seen in absorption with high column densities
and very low emission from young stars \citep{rafelski11}. We suggest that
GC formation sites are brighter than this.

A variety of models for triggering GC formation have been considered, such
as collisions between dwarf galaxies \citep{muratov10} or pressures from a
nearby galaxy \citep{gray11}. Here we discussed GC formation in small
galaxies regardless of their proximity to large galaxies and regardless of
what triggers the star formation. If LAE galaxies are the formation sites,
then any correlation between LAEs and large galaxies in the early universe
could reveal the extent to which metal-poor GCs formed in halo sub-clumps
versus the field. The observation of isolated LAEs would imply that their
metal-poor GCs entered today's galaxy halos later, presumably making the
observed remnant stellar streams by tidal disruption.

Observations show overdensities of LAE galaxies only on the scale of galaxy
clusters, namely, $\sim10^{14}\;M_\odot$ and several Mpc
\citep{kurk04,venemans05,Kovac07, Gawiser07,Tilvi09, overzier06,overzier08}.
This cluster overdensity allows centrally-located elliptical galaxies to
accrete a higher proportion of metal-poor GCs from former LAEs than spiral
galaxies can. Such enhanced accretion is consistent with a greater specific
frequency of metal-poor GCs in ellipticals compared to spirals.

For dwarf galaxy hosts in general, metal-poor GCs should form over a range
of times because their abundance reflects the metal-poor environment of the
dwarf galaxy, rather than the average metallicity of the universe.  NGC 1569
\citep{kobulnicky97} and NGC 1705 \citep{lee04} are examples of moderately
metal-poor environments that make massive clusters today.  They did not fall
into the halos of more massive spiral galaxies, and so their stellar masses
and heavy element abundances continued to increase. Other models have been
concerned with truncating the formation of GCs after a short formation epoch
in order to keep the metallicity low \citep[e.g.,][]{bekki08}. A key
observation will be the age spread of metal-poor GCs. Extended formation
times for metal-poor GCs were recently observed for S0 galaxies by
\cite{chies-santos11}. There was also a suggestion of extended metal-poor GC
formation in one of the stellar population fits by \cite{mendel07} for Milky
Way GCs.

\section{Conclusions}

Metal-poor GCs could have entered the halos of spiral galaxies in the form
of starburst remnants in old dwarf galaxies.  Their low metallicities are
not the exclusive result of an earlier birth time compared to metal-rich
disk and bulge GCs, but rather the result of a lower mass host, considering
the mass-metallicity relation in galaxies.  Based on observations of Milky
Way star-formation and gas fractions for high redshift galaxies, we propose
that a typical metal-poor GC with an original mass of $2\times10^6\;M_\odot$
formed among a total cluster mass of $\sim3\times10^7\;M_\odot$ along with
several $\times10^9\;M_\odot$ of dense gas and $\sim10^7$ to
$\sim10^9\;M_\odot$ of older stars. During a $\sim100$ Myr burst phase when
the star formation rate was several $M_\odot$ yr$^{-1}$, such galaxies would
have AB magnitudes of $\sim26$ and $\sim29$ at wavelengths of 1.4$\mu$m and
3.3$\mu$m for $z\sim2$ and $z=6$, respectively, and the GCs would have
magnitudes of $\sim34$ to $\sim36$ in this redshift range. The host dwarf
galaxies would be $\sim1$ kpc in size and subtend an angle of
$\sim0.15^{\prime\prime}$.

Lyman-$\alpha$ emitting galaxies are plausible formation sites for
metal-poor GCs.  They have the right size, luminosity, star formation rate,
and metallicity. To determine whether LAEs also have the right space density
to form all of today's metal-poor GCs, we integrated the LAE luminosity
function above $\sim2.5\times10^{41}$ ergs s$^{-1}$, which is the expected
Lyman $\alpha$ escape luminosity of a star-forming region that produces a
typical GC today. This limit assumes an average 25\% Lyman-$\alpha$ escape
fraction and one magnitude of extinction for Lyman continuum radiation. The
result equals the space density of today's metal-poor GCs when corrected for
the fraction of LAE galaxies that are in the active phase ($\sim0.4$\%).

JER and SM acknowledge support from NSF AST-0808165. Helpful comments by the
referee are gratefully acknowledged.

\end{document}